\documentclass[nofootinbib,amsmath,12pt]{revtex4}

\newcommand{\beq}{\begin{equation}}
\newcommand{\eeq}{\end{equation}}

\newcommand{\beqa}{\begin{eqnarray}}
\newcommand{\eeqa}{\end{eqnarray}}

\newcommand{\bean}{\begin{eqnarray*}}
\newcommand{\eean}{\end{eqnarray*}}
\newcommand{\ra}{\rightarrow}
\newcommand{\da}{\dagger}
\newcommand{\pa}{\partial}
\newcommand{\mc}[1]{{\mathcal{{#1}}}}

\newcommand{\nn}{\nonumber}

\usepackage{mathbbol}
\usepackage{amsmath}
\usepackage{amssymb}
\usepackage{dcolumn}

\begin{document}

\title{Constrained Mechanics and Noiseless Subsystems}
\author{Tomasz Konopka} \email{tkonopka@perimeterinstitute.ca}
\author{Fotini Markopoulou} \email{fotini@perimeterinstitute.ca}
\affiliation{University of Waterloo, Waterloo, ON N2L 3G1, 
Canada, and \\ Perimeter Institute for Theoretical Physics, 
Waterloo, ON N2J 2W9, Canada}

\begin{abstract}{
Many theories are formulated as constrained systems. We provide a 
mechanism that explains the origin of physical states of a 
constrained system by a process of selection of noiseless 
subsystems when the system is coupled to an external environment. 
Effectively, physical states that solve all the constraints are 
selected by a passive error correction scheme which has been 
developed in the context of quantum information processing. We 
apply this mechanism to several constrained theories including 
the relativistic particle, electromagnetism, and quantum gravity, 
and discuss some interesting (and speculative) implications on 
the problem of time and the status of symmetries in 
nature.}\end{abstract}

\maketitle

\section{Introduction}

Quantum information processing is a relatively young field of 
research compared to, for example, the study of constrained 
systems in classical and quantum mechanics. The amount of 
research in this new field is largely due to its possible 
exciting applications in computation and communication 
\cite{NielsenChuang}. There are also indications, however, that 
some quantum information processing techniques may inspire new 
ways to look at some old problems such as quantum gravity 
\cite{Markopoulou}. In particular, the framework of noiseless 
subsystems has been argued to provide an intuitive picture for 
the emergence of particles and space-time from a fully quantum 
system \cite{Kribs}. Here, we show that the framework of 
noiseless subsystem can provide a basis for thinking about any 
(quantum) mechanical system that is subject to constraints. 

The framework of noiseless subsystems has been developed as a 
tool to preserve fragile quantum information against decoherence 
\cite{DFcomputation}. In brief, when a quantum register (a 
Hilbert space) is subjected to decoherence due to an interaction 
with an external and uncontrollable environment, information 
stored in the register is, in general, degraded. It has been 
shown that when the source of decoherence exhibits some 
symmetries, certain subsystems of the quantum register are 
unaffected by the interactions with the environment and are thus 
noiseless. These noiseless subsystems are therefore very natural 
and robust tools that can be used for processing quantum 
information. 

The outline of the paper is as follows. In the next section, we 
discuss noiseless subsystems in more detail and argue that the 
algebraic structures used in their descriptions are very similar 
to the structures familiar from the study of constrained quantum 
systems. The technical result of this section is the construction 
of a map from a constrained system to the quantum theory of a new 
system coupled to an environment; physical states of the 
constrained system correspond to noiseless subsystems in the new 
theory. Section \ref{s_examples} applies this map to several 
well-known constrained systems, leading to new and sometimes 
speculative interpretations on the role and the origin of 
symmetries such as gauge invariance and time re-parametrization 
invariance. Each example brings out or emphasizes different 
aspects of the noiseless subsystem picture. We end with a 
discussion of the results in section \ref{s_discussion}.

\section{Emergence of Symmetries}

Our goal in this section is to review the formalism of noiseless 
subsystems \cite{DFcomputation} and to compare it with Dirac's 
method for dealing with constrained mechanics. A new picture will 
appear in which solutions to constrained dynamics problems can be 
thought of as noiseless subparts of some quantum 
information-theoretic system.

\subsection*{Noiseless Subsystems}
We consider a quantum theory of a system $S$ coupled to an 
environment (bath) $B$. The full Hilbert space is given by the 
tensor product \beq \nn \mc{H}_{full} = \mc{H}_S \otimes 
\mc{H}_B. \eeq We assume that the system and bath Hilbert spaces 
are compact and discrete. The full Hamiltonian describing 
evolution is \beq \nn \label{Hfull} H_{full} = H_S\otimes I_B + 
I_S\otimes H_B + H_I, \eeq where $H_S$ and $H_B$ are operators 
acting on the system and bath, $I_S$ and $I_B$ are unit operators 
on $S$ and $B$, and $H_I$ encodes the interactions between $S$ 
and $B$. This last term can generically be decomposed as \beq 
\label{Hint} H_I = \sum_\alpha N_\alpha \otimes B_\alpha, \eeq 
where $N_\alpha$ and $B_\alpha$ are a set of operators acting 
only on the system and bath respectively. It is worth pointing 
out that the terms $H_S\otimes I_B$ and $I_S\otimes H_B$ that are 
singled out in $H_{full}$ are only special cases of 
interactions and could in principle be included in the expansion 
of $H_I$.

Operators $H_S,$ $I_S$ and $N_\alpha$, the parts of the 
Hamiltonian that act on the system, generate an algebra which is 
usually called $\mc{A}$. The interpretation of $\mc{A}$ is that 
it comprises all the possible operations (as part of the system 
or the interaction Hamiltonians) that change the state of the 
system. It follows from the fact the Hamiltonian is hermitian 
that $\mc{A}$ is a$\;^\da$-closed, unital algebra. It can be 
decomposed as \beq \nn \mc{A} = \bigoplus_J I_{n_J}\otimes 
\mc{M}_{d_J}, \eeq where the tensor sum is over independent 
algebras $\mc{M}_{d_J}$ of $d_J\times d_J$ matrices, each 
occurring with multiplicity $n_J$. Following through, the Hilbert 
space of the system can similarly be decomposed as \beq \nn 
\mc{H}_S = \bigoplus_J C_{n_J} \otimes C_{d_J}. \eeq An operator 
$N_\alpha \in \mc{A}$ acts on $|a\,b\rangle$, where $a$ and $b$ 
denote states according to the above decomposition, to give \beq 
\label{Nrot} N_\alpha |a\,b\rangle = 
 \sum_{b^\prime} M^\alpha_{b\, b^\prime} |a\,b^\prime\rangle. \eeq 
The matrix $M$ is a rotation of the states labelled by $b$ and 
does not depend on the label $a$. One sees that the subspaces 
labelled by states $a$ in $C_{n_J}$ are acted upon with the unit 
operator by elements of $\mc{A}$ so that they are therefore left 
unchanged during evolution - these subspaces are said to be 
`noiseless' or `decoherence free.' 

From another point of view, consider the density matrix 
$\rho=|a\,b\rangle\langle a\, b|$. The action of the operators 
$N_\alpha$ on $\rho$ is \beq \nn N_\alpha \rho N_\alpha^\da = 
\sum_{b^\prime}\sum_{b^{\prime\prime}} 
M_{b,b^\prime}^\alpha|a\,b^\prime\rangle\langle a\, 
b^{\prime\prime}|M_{b,b^{\prime\prime}}^{\alpha\,\da}. \eeq 
Tracing out the subsystem spanned by the $|b\rangle$ states gives 
\beq \label{tracingB} \mathrm{Tr}_B N_\alpha \rho N_\alpha^\da = 
|a\rangle \langle a|, \eeq which is also equal to $\mathrm{Tr}_B 
\rho$, the partial trace of the original density matrix. This 
further shows that the subspace spanned by states $|a\rangle$ 
(density matrices $|a\rangle\langle a|$) is left invariant by the 
noise operations. Note that if a density matrix $\rho$ is not in 
the special product form $|a\, b\rangle\langle a\, b|$, then 
$\mathrm{Tr}_B N_\alpha \rho N_\alpha^\da \neq \mathrm{Tr}_B 
\rho$, giving the appearance of non-unitary evolution. 

There is an interesting specialization of noiseless subsystems in 
the event where the algebra $\mc{A}$ decomposes so that all the 
matrix algebras $\mc{M}_{d_J}$ are one-dimensional. In this case, 
the form of operators $N_\alpha$ is \beq \label{Nrot2} 
N_\alpha|a,b\rangle = p_{\alpha\, b}|a\,b\rangle, \eeq where the 
phases $p_{\alpha\,b}$ replace the rotations 
$M_{b\,b^\prime}^{\alpha}$ of (\ref{Nrot}). The phases can be 
avoided by using the density matrix formalism, \beq \nn 
N_\alpha|a,b\rangle\langle a,b|N_\alpha^\da = p_{\alpha\, 
b}|a,b\rangle\langle a,b| p_{\alpha\,b}^{\da} = 
|a\,b\rangle\langle a\,b|. \eeq In such special cases, the 
operators $N_\alpha$ are called `stabilizer' elements and the 
invariance is apparent without having to trace out a particular 
subsystem. 

Now, recall that the system Hilbert space is coupled to an 
environment. Thus a full state can be written as $\rho=|\psi\, 
\phi\rangle\langle \psi\,\phi|$ where $\psi$ is a state of the 
system Hilbert space and $\phi$ is a state of the environment. 
Evolution is generated by the unitary operator 
$U_{full}=\exp(i\tau H_{full})$. In time $\tau$, dropping the 
subscripts, the density matrix changes \beq \nn \rho \ra U\rho 
U^\da \sim \rho + \frac{\tau^2}{2}\left(2H\rho H - H^2\rho 
 - \rho H^2 \right),\eeq the approximation being made in the 
limit of short evolution times. For the case of the interaction 
Hamiltonian (\ref{Hint}), consider the operators $N$ to act as 
$N|\psi\rangle = |\psi^\prime\rangle$ and $N|\psi^\prime\rangle = 
|\psi^{\prime\prime}\rangle.$ Then, after a short evolution and 
after tracing out states of the environment, one has \beq \nn 
|\psi\rangle\langle\psi| \ra |\psi\rangle\langle\psi| + 
\frac{\tau^2}{2} \left(2|\psi^\prime\rangle\langle\psi^\prime| - 
|\psi^{\prime\prime}\rangle\langle\psi| - 
|\psi\rangle\langle\psi^{\prime\prime}|\right).\eeq 

The nature of the resulting state depends on the type of initial 
state that we start with. In the special case where the original 
state is noiseless and obeys (\ref{Nrot2}) with unit phase, i.e. 
$|\psi^\prime\rangle = |\psi^{\prime\prime}\rangle=|\psi\rangle$, 
the terms in the parenthesis cancel and the evolution is trivial. 
In other cases, the interpretation of the resulting state is less 
self-evident. For example, when the original state $|\psi\rangle$ 
is noiseless and obeys (\ref{Nrot}), the evolution appears to be 
trivial only after tracing out a subsystem similarly as in 
(\ref{tracingB}). A different situation arises when the original 
state does not satisfy any of the noiselessness conditions. Then, 
the result of the evolution is a state that is, generically, 
mixed and whose `extent' of mixture grows as the evolution 
progresses (the new terms in density matrix are proportional to 
$\tau$). Thus, from the point of view of the system, noisy states 
evolve according to a dissipative, non-unitary, dynamics.  

Decoherence-free states are of importance for quantum information 
processing because they can be used to reliably store information 
for long periods of time. For information processing, however, it 
is also very important to be able to manipulate or change 
information in order to perform computations. To this end, it is 
interesting to define the possible operations that can be applied 
to a noiseless states without ruining its noiseless feature. 
These operations are elements of the algebra  $\mc{A}^\prime$ of 
all elements that commute with the interaction algebra $\mc{A}$. 
That is, an operator $A^\prime\in\mc{A}^\prime$ can be used to 
manipulate a noiseless state if and only if $[A,A^\prime]=0$ for 
all $A\in \mc{A}$. For more details on identifying noiseless 
subsystems via the commutant of the algebra of operators on the 
system see \cite{DFcomputation}.

\subsection*{Constrained Systems}

Shifting slightly, we now briefly review the standard method of 
dealing with quantum constrained systems (see, for example,
 \cite{QGreviews}). The Hilbert space of 
an unconstrained system is called the kinematical Hilbert space 
$\mc{H}_{kin}$. Constraints are represented by operators $C_a$ 
that form a closed, first-class algebra 
$[C_a, C_b] = f_{ab}^{\;\;\;c}C_c$ for some structure constants 
$f_{ab}^{\;\;\;c}$. Physical states of the system are defined to 
be those that satisfy the constraint equations $C_a 
|\psi\rangle_{phys} = 0;$ the span of these states forms the 
physical Hilbert space, $\mc{H}_{phys}$. An important aspect of 
understanding constrained systems is the construction of the 
algebra $\mc{D}$ of Dirac observables. Operators in this algebra 
commute with the constraints and thus measure physical 
(invariant) properties of physical states. In other words, $D$ is 
an observable, $D \in \mc{D}$, if and only if $[D,C_a]=0$. 

There are significant similarities in the algebraic structures 
that are relevant to the constrained systems and that appear in 
the discussion of noiseless subsystems. Specifically, in each 
case one has two distinct algebras that commute. The aim of the 
present work is to probe this similarity and establish a 
connection between constrained systems and noiseless subsystems. 
This is accomplished by constructing a mapping from a constrained 
system to a noiseless subspace. 

Consider a system subject to a set of first-class constraints 
$C_a$. Consider also an identity operator $I$ on the kinematical 
Hilbert space $\mc{H}_{kin}$ and define new operators $N_{a\, 
\lambda}=(I+\lambda C_a)$. Then if $C_a|\psi\rangle_{phys} = 0$, 
operators $N_{a\, \lambda}$ stabilize physical states for all 
$\lambda$, $N_{a\, \lambda}|\psi\rangle_{phys} = 
|\psi\rangle_{phys}$. Thus, an alternative description of the 
constrained system starts to develop in which $\mc{H}_{kin}$ can 
be identified with $\mc{H}_{S}$ and the new stabilizer elements 
$N_{a\, \lambda}$ generate the algebra $\mc{A}$. Recall that 
elements of $\mc{A}$ have the interpretation of being operations 
that couple the system to an environment. Thus, this approach 
suggests $\mc{H}_{kin}$ should be coupled to a new Hilbert space 
$\mc{H}_B$ representing an environment or bath. 

The interaction Hamiltonian (\ref{Hint}) for the constrained 
system and environment will have the form
 \beq H_I = \sum_{a} N_a 
\otimes B_a = \sum_a \left( 1 \otimes B_a + \lambda C_a \otimes 
B_a \right),\eeq for some operators $B_a$ acting on the 
environment. Incidentally, the decomposition of the interaction 
terms on the right hand side make the first term appear as 
operators acting on the environment only, i.e. being part of 
$1\otimes H_B$ of (\ref{Hfull}). Only the terms proportional to 
the constraints are therefore part of the `true' interaction 
Hamiltonian, \beq \label{newHint} H_I \ra \sum_a C_a \otimes B_a. 
\eeq In short, what we now have is a new quantum system with a 
full Hilbert space $\mc{H}_{full}=\mc{H}_{kin}\otimes \mc{H}_B$ 
governed by a Hamiltonian of the form (\ref{Hint}) with $H_S$ 
given by the Hamiltonian of the constrained problem, $H_B$ given 
by the operators $B_a$, and $H_I$ given by (\ref{newHint}). 

The noiseless states of this new theory are, by construction, 
solutions to the constraints $C_a$ that we started with. They 
therefore exhibit all the physical properties that the solutions 
to the constrained problem do. Since the environment in the 
quantum information theoretic description is not really of 
interest from the point of view of the constrained dynamics 
problem, it should be traced out. As a result, the noiseless 
states evolve unitarily under the full Hamiltonian while the 
noisy states, which do not satisfy the constraint equations, 
decay non-unitarily and as such are not of physical interest. 

The commutant $\mc{A}^\prime$ in the noiseless subspace picture 
is the set of all operators that commute with the constraints 
$C_a$. Thus, there is also a close correspondence between 
$\mc{A}^\prime$ and $\mc{D}$, up to the status of the unit 
operator. The unit is always a Dirac observable and is thus in 
$\mc{D}$. On the noiseless subsystem side, however, the unit 
operator is also included in the algebra $\mc{A}$ (recall that 
$\mc{A}$ is assumed unital). The interesting correspondence, 
therefore, should be made between the non-trivial elements of 
$\mc{A}^\prime$ and the non-trivial Dirac observables.

\section{Examples \label{s_examples}}

The transition from constrained dynamics to noiseless subsystems 
is developed further in the following series of examples, 
starting from a straight-forward non-relativistic particle 
subject to a linear constraint, through gauge theory and the 
relativistic particle, and culminating in a discussion of quantum 
gravity. 

\subsection*{Momentum Constraint}

A very simple example of a constrained system is a classical
non-relativistic particle moving in two spatial dimensions $x$ 
and $y$ under the restriction that its $y$ momentum be zero. The 
particle is described by a Hamiltonian \beq \nn H = \left( 
\frac{p_x^2}{2m} + \frac{p_y^2}{2m} + \beta C\right) \eeq that 
consists of kinetic terms and the constraint $C$ with its 
Lagrange multiplier $\beta$. For this example, $C=p_y$. The 
discussion below of the quantum version is only an outline but 
fully conveys the spirit of the mapping from constrained dynamics 
problems to noiseless subsystems. 

The kinematical Hilbert space for the particle is taken to be the 
momentum space where states are labelled $|\psi\rangle = |p_x, 
p_y\rangle$. The constraint requires that $p_y|p_x,p_y\rangle = 
0$. Intuitively, the states that make up the physical Hilbert 
space are $|p_x,0\rangle$. Now consider the description of the 
particle motion from the noiseless subspace point of view. In 
this view, the full system is composed of the particle coupled to 
an environment. We define an identity operator $I$ such that 
$I|\psi\rangle = |\psi\rangle$ for all $|\psi\rangle$. Then, if a 
state satisfies $p_y|\psi\rangle = 0$ it also satisfies $(I+p_y 
\lambda)|\psi\rangle = |\psi\rangle$ for all $\lambda$. Thus, 
$N_\lambda = (I+p_y \lambda)$ is a stabilizer of physical states. 
(Of course, the operator $I$ by itself is also a stabilizer of 
physical states, but it is not interesting since it also 
stabilizes non-physical states.) 

According to the noiseless subsystems approach, the particle is 
coupled to an environment. The full system of the particle and 
environment evolve under the Hamiltonian \beq \nn H_{full} = 
\left( \frac{p_x^2}{2m}+\frac{p_y^2}{2m}\right) \otimes 1 + 
1\otimes B + p_y \otimes B, \eeq where $B$ is some unspecified 
unitary operator that describes the evolution of the environment 
as well as the coupling of the particle to the environment. It is 
interesting to note that the states $|p_x, 0\rangle$ are not the 
only ones that can be considered as noiseless. Other momentum 
eigenstates, such as $|p_x, p_y\rangle$ with arbitrary $p_y$ can 
also be used. Operators $N_\lambda$ act on such states as \beq 
\nn N_\lambda|p_x, p_y\rangle = |p_x\rangle \otimes 
(1+P_y)|p_y\rangle = (1+p_y)|p_x, p_y\rangle,\eeq which is a 
(unnormalized) form of of (\ref{Nrot2}) with non-trivial phase. 
However, superpositions of states with different $p_y$ do not 
satisfy the stabilizer condition and would thus appear to be 
noisy in evolution. Thus the system can be viewed as containing 
multiple (indeed, an infinite number of) noiseless subspaces, one 
for each value of $p_y$. The equal status of all the noiseless 
subspaces can be traced back to translational symmetry in the 
original constrained system picture, i.e. the freedom to rewrite 
the constraint from $p_y=0$ in one reference frame to another 
value of $p_y$ in some other reference frame. 

In both the constrained dynamics picture and the noiseless 
subspace picture, operators $D$ such that $[D,C]=0$ are 
important. In the constrained dynamics picture, such operators 
are called Dirac observables. For example, $p_x$ and $x$ can be 
Dirac observables. In the noiseless subsystem picture, the 
operators $D$ can be used to manipulate the information stored in 
the particle state. For example, the operator $p_x$ can be used 
to read off the $x$ momentum in an eigenstate. As another 
example, an operator $x$ can be used to shift the momentum of the 
particle by a certain amount. It should be stressed that these 
operators could not be used reliably on noisy states.

An attractive feature of this simple example is that the role of 
the environment, which is fundamental to the description of the 
particle in the noiseless subsystem picture, can also be 
understood from the perspective of the constrained system. There, 
the Lagrange multiplier $\beta$ is interpreted as a force that 
determines the particle's momentum. Since the source of the force 
is external to the particle, the constrained dynamics description 
of the particle also implicitly makes use of an environment. In 
some sense, then, the noiseless subsystem worldview can be seen 
to emphasize a feature that is present but that is often 
overlooked in the constrained mechanics formalism. These external 
forces, usually hidden in the constrained dynamics framework, are 
brought to the fore in the quantum information theoretic 
description of the system in terms of noiseless subsystems. In 
fact, the strategy for describing solutions to constrained 
problems in terms of noiseless subsystems will be to 
introduce/postulate a new environment system for every Lagrange 
multiplier and write down suitable interactions to generate the 
desired noiseless dynamics.

\subsection*{Gauge Theory}

As another example, we consider electromagnetism with the action 
\beq \nn \label{Sem} S = -\frac{1}{4} \int d^4x \; 
F_{\mu\nu}F^{\mu\nu}, \eeq where 
$F_{\mu\nu}=\pa_{\mu}A_{\nu}-\pa_\nu A_\mu$. The action is 
invariant under gauge transformations $A_\mu \ra A_\mu + \pa_\mu 
\alpha$ where $\alpha$ is any space-time function. The momenta 
that are conjugate to $A_\mu$ are $\pi^0=0$ and $\pi^i = F^{0i}$. 
The Hamiltonian for the theory is \beq \nn H = \int d^3x \; 
\left(\frac{1}{2}\pi_i \pi^i + \frac{1}{2}F_{ij}F^{ij} - A_0 
\pa_i \pi^i  \right). \eeq  In the last term, $A_0$ appears as a 
Lagrange multiplier imposing the constraint \beq \nn C = \pa_i 
\pi^i = \pa_i \pa^i A_0 - \pa_i \pa_0 A^i. \eeq The effect of the 
constraint is to reduce the number of physical, propagating 
degrees of freedom in the vector potential down to two, giving 
electromagnetism an interpretation as a theory of massless spin 
one particles (photons) propagating at the speed of light.

Since electromagnetism is a theory of a four-dimensional vector 
field, the Hamiltonian should be a function of all four 
components of $A_\mu$ and their conjugate momenta, 
$H=H(A_0,p_0,A_i,p_i)$. The Hamiltonian above, however, is of the 
special form \beq \nn H = H_S(\pi_i, A_i) - C(\pi^i, A^i) A_0 
\eeq whereby $H_S$ does not depend on the scalar potential $A_0$ 
nor its momentum, and the second term is clearly split into two 
factors, one of which depends $A_0$ and another one that does 
not. This kind of splitting suggests writing the Hilbert space 
of electromagnetism as $\mc{H} = \mc{H}_S \otimes \mc{H}_B$, 
treating the the vector potential $A_i$ as the `system' and the 
scalar potential $A_0$ as the `bath.' The Hamiltonian is 
therefore composed of a piece $H_S\otimes 1$  that evolves the system only 
and an interaction term $H_I = C \otimes A_0$. 

Drawing on the earlier general discussion, we would like the 
system part of the interaction term to act as a stabilizer on 
physical states. At the moment, that part of $H_I$ annihilates 
physical states. To make electromagnetism match the noiseless 
subsystems scheme, we replace the existing interaction term by 
$H_I \ra N \otimes A_0$ where $N=1+C$. The effect of this 
exchange is to introduce a new term into the Hamiltonian: 
\beq \nn H = H_S(\pi_i, A_i) - C(\pi^i, A^i) A_0 - A_0.\eeq Now
the noiseless states in the system part, after tracing out the 
scalar potential degrees of freedom, should be exactly the ones 
that correspond to the transverse polarizations of photons. 

The physical picture that emerges from the noiseless subsystem 
framework is as follows. The full set of fields ($A_0, A_i$) 
evolves according to a well defined Hamiltonian which treats the 
vector potential as the `system' and the scalar potential as the 
environment. Due to the interaction term, only certain states of 
vector potential are noiseless. So only those states can be 
expected to be preserved in the `long term' and be 
seen/observed/detected in the laboratory - photons are 
interpreted as excitations in the noiseless sector of the vector 
potential. Having experience dealing only with the noiseless 
states, we are usually inclined to describe the experimental 
results using a theory that is re-parametrization invariant, i.e. 
a gauge theory. However, if experiments could give access to the 
full spectrum of states as opposed only to the noise-free ones, 
in effect allowing us to measure the scalar potential directly, 
then the gauge invariance would not be a true symmetry of the 
full system. This is reminiscent of the `random dynamics' 
research program where gauge symmetry is emergent
\cite{Nielsenchaos}.

The extra term in the Hamiltonian has a similar effect to 
introducing a gauge fixing condition. This is not a problem in 
the present sense because, at the same time, we postulate that 
the scalar potential is not observable and focus the discussion 
on noiseless states in the vector potential Hilbert space; 
tracing out the scalar potential gets rid of the rigidity of the 
fixed gauge. At the end, the quantum information theoretic 
description ends up with the same physical solutions as the 
original gauge theory. In this sense we can say that gauge 
invariance is an emergent property in the noiseless states.

\subsection*{Relativistic Particle}

Another familiar example of a constrained system is the 
relativistic particle moving in a flat background (metric 
$\eta_{\mu\nu}=diag(-1,1,1,1)$). Its Lagrangian is given by \beq 
\nn L = -m\sqrt{-\eta_{\mu\nu} \dot{q}^\mu \dot{q}^\nu}. \eeq The 
overdot denotes time derivatives $\pa/\pa_\tau$. The conjugate 
momenta to $q_\mu$ are $p_\mu = (\pa L)/(\pa \dot{q}^\mu) = (m^2 
\dot{q}_\mu)/(L), $ and the standard Hamiltonian $H = p_\mu 
\dot{q}^\mu -L = 0$ vanishes due to time-reparametrization 
invariance of the Lagrangian. The system is instead characterized 
by the constraint \beq \nn C = -p_0^2 + p_i^2 +  m^2,\eeq which 
puts the relativistic particle on-shell. The total Hamiltonian 
therefore consists of only this constraint, \beq \nn H = \beta C, 
\eeq where $\beta$ is a Lagrange multiplier. To obtain the 
physical states of the particle, consider first $\mc{H}_{kin}$ to 
be the space of wavefunction of four momentum variables, spanned 
by the states $|p_0, p_i\rangle$. Due to the constraint, only 
three of the four momenta can be independent. It is convenient to 
chose $p_0$ as the dependent variable. States \beq \nn 
|\psi\rangle_{phys}=|p_{0\, phys}, p_i\rangle\qquad\; p_{0\, 
phys}=\sqrt{p_i^2 +m^2} \eeq satisfy the constraint, 
$C|\psi\rangle_{phys} =0$.

To view the relativistic particle from a noiseless subsystem 
point of view, we consider the algebra $\mc{A}$ generated by the 
constraint $C$ and the identity operator. We define operators 
$N_\lambda = 1 +\lambda C$ acting on the system Hilbert space 
spanned by states labelled by four-momenta. The full Hamiltonian 
describing the evolution of the particle coupled to the 
environment is defined as \beq \nn \label{Hparticlefull} H_{full} 
= 1\otimes B + C \otimes B, \eeq in analogy to (\ref{Hfull}) but 
with $H_S=0$ due to the actual Hamiltonian of the relativistic 
particle being zero in the constrained dynamics picture. The 
particle and the environment evolve in the usual way via an 
evolution operator $U=\exp(i\tau\, H_{full})$; the evolution is 
naturally parametrized by a new external time variable $\tau$. 

In the quantum information theoretic picture, a generic initial 
state evolves into a totally mixed background with particle-like 
excitations. This situation can be compared to the description of 
signals in liquid-state Nuclear Magnetic Resonance experiments. 
There, a liquid sample is viewed as consisting of many small 
randomly-oriented spins. When a low-frequency pulse is applied, 
the spins tilt slightly with the effect of generating a net 
magnetization in the sample. An effective density matrix can be 
used to describe the state of the sample which is composed of a 
sum of a total mixed state and a small particle contribution. 
Remarkably, the particle contribution actually behaves like a 
real particle and can even be successfully employed in 
experimental quantum information processing \cite{NMRstuff}. The 
proposal here is to view the relativistic particle in a similar 
manner - as an excitation over a noisy background.

Note that noiseless states evolve as if the Hamiltonian were zero 
exactly as in the original constrained system. Thus, these states 
exhibit an emergent time-reparametrization invariance property. 
However, in the noiseless subsystems picture, the `true' 
Hamiltonian is actually $H_{full}$ and is nonzero. There is no 
`problem of time' as the evolution of the environment provides a 
well defined clock to measure time flow by. This is another novel 
feature introduced by the noiseless subsystems viewpoint, and it 
may be helpful in reducing the conceptual difficulties that arise 
in the study of time re-parametrization invariant systems. 

The proposed viewpoint, in a sense, is orthogonal to the much 
discussed relational approach (see for example 
\cite{Poulin:2005dn,time}) where the introduction of a background 
time is seen as something that should be avoided. Of particular 
interest is the work of Poulin \cite{Poulin:2005dn} who uses 
noiseless subsystems and quantum information theoretic tools to 
write a relational formulation of quantum theory that is 
originally expressed in terms of a background time. In contrast, 
we argue in the reverse direction that the relational features 
usually ascribed to physical systems such as the relativistic 
particle can be understood as arising out of a non-relational 
theory of the system under consideration coupled to an 
environment. 

Relationalism can be restored, however, by considering this 
non-relational theory as part of another, larger relational 
theory. Then, density matrices form a hierarchy \beq \nn 
\rho_{rel} \hookrightarrow \rho_{non\,rel} \hookrightarrow 
\rho_{new\,rel} \eeq where $\rho_{rel}$, $\rho_{non\,rel}$ and 
$\rho_{new\,rel}$ describe, respectively, the usual relativistic 
particle, the particle together with the environment, and the 
particle together with an environment as well as another 
auxiliary system (a clock). The transition between a density 
matrix in a large Hilbert space to a density matrix in a smaller 
space is performed by tracing out the redundant degrees of 
freedom. The bottom line is that the fixed background structure 
of the environment can be treated in a relational manner if 
relationalism is desired.

\subsection*{Quantum Gravity}

As a final example, we consider the quantization of general 
relativity. As is well known, gravity, like the relativistic 
particle,  is a totally constrained system \cite{QGreviews}. The 
Hamiltonian is a sum of first-class constraints, \beq \nn H = 
\int_\Sigma A_0^i G^i + N_a D_a + N D. \eeq The integral is taken 
over a three-dimensional manifold $\Sigma$. The Lagrange 
multiplier $A_0^i$ is of the kind appearing in the gauge theory 
example and implements the Gauss constraint $G^i$. The lapse 
function $N$ that implements the Hamiltonian constraint $D$ is 
akin to the Lagrange multiplier appearing in the relativistic 
particle example. The remaining multipliers $N_a$ that implement 
the three diffeomorphism constraints $D_a$ are characteristic to 
this example, but they can be treated using similar methods. 

Multiple constraints in the gravity Hamiltonian can be treated in 
sequence and on an individual basis. By this we mean that, in 
general, each constraint (suppose there are $n$ of them) can have 
a separate environment Hilbert space $\mc{H}_{Bn}$ associated 
with it giving $\mc{H}_{full} = \mc{H}_S\otimes 
\mc{H}_{B1}\otimes \cdots \otimes \mc{H}_{Bn}$. Solutions to the 
$n$-th constraint can be found in the space $\mc{H}_S\otimes 
\mc{H}_{B1}\otimes \cdots \otimes \mc{H}_{B(n-1)}$ (or its dual) 
as the noiseless states with respect to the appropriate coupling. 
After having characterized the solutions/noiseless states of this 
constraint, another constraint can be considered to further 
restrict the set of states that are of physical significance, and 
so on until all the constraints are taken care of. At each step, 
the size of the Hilbert space decreases until one finally 
determines the noiseless states in the original system Hilbert 
space $\mc{H}_S$.

If we are interested in the exact solutions of the constraints, 
it is of no significance whether the characterization of the 
solutions is done via standard methods or via quantum information 
theoretic tools. In particular, we can simplify our discussion of 
quantum gravity by using the well-known result that states 
invariant under gauge transformations and spatial 
diffeomeorphisms can be labelled by spin networks 
\cite{Ashtekar}. To formally obtain solutions to full quantum 
general relativity, then, we should couple the spin-networks to 
an environment and define the interaction Hamiltonian in terms of 
the Hamiltonian constraint. The noiseless spin networks in this 
scheme are the physical solutions of interest; it is likely that 
these noiseless states would have proper descriptions in terms of 
classical spacetimes. Unfortunately, the quantum information 
theoretic approach does not make the problem of actually writing 
down simple expressions for these states any easier than in the 
standard Dirac quantization program. A perhaps promising feature 
of the noiseless subsystem approach, however, is that these 
physical states should appear dynamically as invariant states out 
of a generic initial state of a system and environment.

Backtracking to the core picture of coupling the kinematical 
Hilbert space of gravity to an environment, observe that the 
symmetries such as gauge-invariance, diffeomeorphism-invariance, 
and time re-parametrization invariance are not fundamental 
features of the full system comprising the various environments. 
In the quantum information theoretic picture, states in the full 
Hilbert space spanned by gravitational and bath degrees of 
freedom act as if they were coupled to a fixed space and time 
background. Thus an observer having access to the full Hilbert 
space can follow the evolution of a gravity state using a set of 
external variables using the methods of standard quantum 
mechanics. It is only the process of ignoring, or tracing out the 
background environment that reproduces the background-independent 
features of general relativity. Tracing out the environment and 
focusing attention on the noiseless states is of course motivated 
by observations of a four-dimensional universe obeying the 
Einstein's equations to a high degree of accuracy. 

Adding an environment to the universe is certainly a strange move 
with interpretational issues if the quantum theory of gravity is 
simply the quantization of the known gravity and matter.  In that 
case the noisy states are unphysical.  However, the situation is 
different in quantum gravity approaches, such as condensed matter 
approaches, in which general relativity is expected to be an 
effective theory describing the behavior of the  low energy 
excitations of an underlying system.  In that case, an 
environment  and usually a true Hamiltonian is already present in 
the fundamental theory and the question is how a constrained 
theory can arise at the effective level.  There are similar 
questions in Causal Dynamical Triangulations, in which the full 
theory has a time parameter.  In this note we wish to suggest 
that general relativity may be the noiseless sector of the 
underlying quantum theory of gravity.

\section{Discussion \label{s_discussion}}

The main result of this work is the connection of physical states 
of constrained systems to noiseless subsystems of quantum systems 
coupled to an environment. An explicit and simple construction is 
provided to map a system subject to first class constraints to 
another system that interacts with an external environment, where 
the interactions provide a mechanism for implementing the 
symmetries of the constrained system as noiseless states. The 
equivalence of the two formalisms can be seen in the fact that 
the algebra of Dirac observables of the constrained system is 
isomorphic to the algebra of non-trivial unitary operations that 
can be performed on the new coupled system.

The relation of constrained systems to noiseless subsystems 
offers a fresh perspective on the status of several important 
symmetries such as gauge invariance in electromagnetism and 
quantum gravity, and time-reparametrization invariance in the 
relativistic particle and quantum gravity. Indeed, the quantum 
information theoretic point of view implies that symmetries of 
physical states may arise, or emerge, dynamically in noiseless 
states. These states are defined by interactions of a system with 
an environment and as a result of ignoring the evolution of that 
environment. In the case of the Gauss constraint (in 
electromagnetism as well as in quantum gravity), the environment 
can be economically thought of as being the scalar potential. In 
the case of other constraints, such as the Hamiltonian constraint 
(in the relativistic particle and quantum gravity), the 
environment should be thought of as a truly external object. In 
all cases, however, symmetries of physical states generated by 
the constraints loose their fundamental status and are in fact 
not present in generic states of the full system including the 
environment.

Emergence of symmetries is a concept that has been already 
studied from various perspectives and discussed in the context of 
quantum gravity. Notably, the notion of emergence is a natural 
one in the context of condensed matter and has been used to study 
many aspects of gravitation \cite{CondMatt} including the role of 
the cosmological constant in quantum gravity 
\cite{Dreyer:2004bc}. A well known result supporting the 
emergence approaches is the equivalence of Euclidean quantum 
theory in the Feynman path integral formulation and statistical 
mechanics. Other indicative works reveal connections between 
general relativity and thermodynamics \cite{Jacobson:1995ab}, and 
quantum features in Bohmian mechanics to the concept of 
equilibrium \cite{Valentini:2004yw}. Discussions of emergent 
features can also be found in the literature on causal sets 
\cite{Sorkin:2003bx}, causal dynamical triangulations \cite{CDT}, 
string-net condensation \cite{Levin}, and quantum information 
theory \cite{Lloyd:2005js,Kribs}. Emergent gauge symmetry is also 
discussed in lattice theories \cite{Nielsenchaos}.

Although the notion of an `environment of the universe' that 
arises in our discussion of quantum gravity may sound slightly 
offbeat, it should be noted that the idea of coupling 
gravitational degrees of freedom to external systems is in fact 
in common use under different names. In quantum cosmology, for 
example, the Hilbert space of an FRW spacetime can be coupled to 
a scalar field to model inflation. Coupling gravity to a scalar 
field is also a useful technique for introducing clock variables 
to be employed in defining Dirac observables for quantum gravity 
\cite{PartialObservables}. The novelty in this work is the 
proposal for a concrete kind of coupling between the 
gravitational and external Hilbert spaces given in terms of the 
stabilizer operators, and the resultant dynamical emergence of 
physical states as noiseless subsystems during evolution. 

The introduction of an environment can be criticized as bringing 
a background to theories such as the relativistic particle or 
gravity that are otherwise thought of as relational and 
background independent. We  can defend the environment against 
this argument in two ways. First, note that the presence of a 
constraint in classical mechanics implies that there is an 
external force acting on the system. Making a straight-forward 
generalization that all constraints arise as a result of an 
interaction with an external source, the presence of an 
environment actually appears quite natural. Second, it is known 
that a non-relational theory can be mapped into a relational one, 
for example by enlarging the configuration space by some clock 
variables. Thus, the fixed-background formulation of the 
environment can in principle be generalized to obtain a 
relational version of the noiseless subsystems-based theory. The 
above criticism, therefore, is not a fundamental obstacle and it 
should not deter us from considering theories in which symmetries 
arise dynamically out of interactions with a fixed structure.  On 
the contrary, this formalism is suggestive of the existence of 
recently proposed dualities between background-independent and 
background-dependent theories \cite{Husain:2005xj}. An important 
benefit is the resulting true Hamiltonian instead of the 
constraint, a feature that is expected to bring the low-energy 
problem for quantum gravity theories to a level similar to that 
of ordinary condensed matter systems.  

\vspace{0.3cm} {\bf Acknowledgements.$\;\;$} We would like to 
thank Lee Smolin, particularly for his comments regarding gauge 
invariance.

\end{document}